\documentclass[a4paper,11pt]{article}

\newcommand{\be}{\begin{eqnarray}}
\newcommand{\ee}{\end{eqnarray}}

\usepackage{jheppub} 

\usepackage[T1]{fontenc} 

\title{\boldmath
Monte Carlo study of Lefschetz thimble structure in 
one-dimensional Thirring model at finite density 
}

\author[a]{Hirotsugu Fujii,}
\author[b]{Syo Kamata}
\author[a]{and Yoshio Kikukawa}

\affiliation[a]{Institute of Physics, University of Tokyo, Tokyo 153-8092, Japan}
\affiliation[b]{Department of Physics, Rikkyo University, Tokyo 171-8501, Japan}

\emailAdd{hfujii@phys.c.u-tokyo.ac.jp}
\emailAdd{skamata@rikkyo.ac.jp}
\emailAdd{kikukawa@hep1.c.u-tokyo.ac.jp}

\preprint{
\begin{flushright}
UT-Komaba/15-7\\RUP-15-21
\end{flushright}
}

\abstract{
We consider the one-dimensional massive Thirring model formulated on the lattice 
with staggered fermions and an auxiliary compact vector (link) field, 
which is exactly solvable and shows a phase transition with increasing the chemical potential of fermion number: 
the crossover at a finite temperature and the first order transition at zero temperature. 
We complexify its path-integration on Lefschetz thimbles and examine its phase transition 
by hybrid Monte Carlo simulations on the single dominant thimble. 
We observe a discrepancy between the numerical and exact results in the crossover region
for small inverse coupling $\beta$ and/or large lattice size $L$,  
while they are in good agreement in the lower and higher density regions.
We also observe that the 
discrepancy persists in the continuum limit 
to keep the temperature finite
and it becomes more significant toward the low-temperature limit.
This numerical result is consistent with our analytical study of  the model 
and implies that the contributions of subdominant thimbles should be summed up 
in order to reproduce the first order transition in the low-temperature limit. 
}

\begin{document} 
\maketitle
\flushbottom

\section{Introduction}
%

%

The physics of QCD at finite temperature and density is one of the most important subjects
in high energy physics and also in cosmology and astrophysics. 
To investigate QCD, 
especially its static and thermodynamic properties, 
the Monte Carlo simulation of lattice QCD has proved to be a powerful method. 
However, in the extreme condition of low temperature and high density,  
the sign problem in lattice QCD,  
caused by introducing the baryon-number chemical potential, 
prevents us from the thorough study of the properties of QCD\cite{deForcrand:2010ys}. 
%
%
Recently two alternative approaches to the problem have attracted much attention  
-- 
complex Langevin dynamics\cite{Parisi:1984cs,Klauder:1983zm,Klauder:1983sp}
and Lefschetz thimble method\cite{Witten:2010cx,Witten:2010zr,Pham:1983}.
Both methods are based on the complexification of dynamical field variables.\footnote{
Recent research activities include 
refs.\cite{
Aarts:2008rr,
Aarts:2008wh,
Aarts:2009hn, 
Aarts:2009yj,
Aarts:2009dg,
Aarts:2009uq,
Aarts:2010aq,
Aarts:2010gr,
Aarts:2011ax,
Aarts:2011sf,
Aarts:2011zn,
Seiler:2012wz,
Aarts:2012ft,
Pawlowski:2013pje,
Pawlowski:2013gag,
Aarts:2013bla,
Aarts:2013uxa,
Aarts:2013uza, 
Sexty:2013ica,
Aarts:2013fpa,
Giudice:2013eva,
Mollgaard:2013qra,
Sexty:2013fpt,
Aarts:2013nja,
Bongiovanni:2013nxa,
Aarts:2014nxa,
Aarts:2014bwa, 
Sexty:2014zya,
Sexty:2014dxa,
Bongiovanni:2014rna,
Aarts:2014kja,
Aarts:2014cua,
Aarts:2014fsa,
Mollgaard:2014mga,
Makino:2015ooa,
Aarts:2015oka, 
Nishimura:2015pba, 
Aarts:2015yba,
Nagata:2015uga,
Fodor:2015doa,
Tsutsui:2015tua} for the complex Langevin dynamics and
refs.\cite{
Cristoforetti:2012su,
Cristoforetti:2013wha,
Fujii:2013sra, 
Mukherjee:2014hsa,
DiRenzo:2015foa,
Tanizaki:2014tua,
Kanazawa:2014qma,
Cristoforetti:2014gsa,
Tanizaki:2014xba, 
Tanizaki:2015pua,
Cherman:2014ofa,
Behtash:2015kna, 
Scorzato:2015,
Fukushima:2015qza,
Tanizaki:2015rda,
Fujii:2015bua} 
for the Lefschetz thimble method.
The authors refer the reader to refs.\cite{Sexty:2014dxa,Scorzato:2015} for reviews of  these approaches. 
} 

In our previous work\cite{Fujii:2015bua}, we have applied the Lefschetz thimble method to 
the one-dimensional lattice Thirring model. 
The model is exactly solvable and shows a phase transition with increasing the chemical potential of fermion number, 
the crossover at a finite temperature and the first order transition at zero temperature, 
which is similar to the expected property of QCD. 
%
In this model, 
we have obtained all the critical points and
examined the thimble structure
by inspecting the solutions of the gradient flow equation, 
the values of the action at the critical points 
and the Stokes phenomena. 
And we  have 
identified the set of the thimbles which contribute to the path-integral and
have classified the dominant thimbles  for given parameters, $L$, $\beta$,  $m$ and $\mu$. 
Our result there suggests
that 
one should sum up the contributions of subdominant thimbles 
in order to reproduce the rapid crossover and the first-order transition in the low-temperature limit.


In this article, we consider the same one-dimensional Thirring model at finite density and perform Monte Carlo simulations taking the most dominant  thimble (referred to as
${\cal J}_{\sigma_0}$ in \cite{Fujii:2015bua}) 
with the HMC algorithm proposed in ref.~\cite{Fujii:2013sra}.
%
%
We will examine to what extent the HMC simulation on the single dominant 
thimble ${\cal J}_{\sigma_{0}}$ works for this model 
by comparing our numerical results with the exact ones. 

This paper is organized as follows.
In section~2, we introduce  the one-dimensional lattice Thirring model  
and apply the Lefschetz thimble method to the model.
In section~3, we describe our HMC simulation details and present our numerical results.
Section~4 is devoted to summary and discussion.

\section{One-dim. Lattice Thirring model complexified on Lefschetz thimbles}

In this section, first we introduce a lattice formulation of the one-dimensional massive 
Thirring model\cite{Pawlowski:2013pje,Pawlowski:2014ada} and discuss its property at finite 
temperature and desity.
Next we apply the Lefschetz thimble method to this lattice model.
The method is based on the complexification of the field variables and the 
decomposition of the original path-integration contour into the cycles called Lefschetz thimbles.
See refs.~\cite{Witten:2010cx,Cristoforetti:2012su,Fujii:2013sra} for the detail of the 
approach and ref.~\cite{Fujii:2015bua} for the detail of the Lefschetz thimble structure of the Thirring model

\subsection{One-dimensional massive Thirring model on the lattice}
\label{sec:def_action}

The one-dimensional lattice Thirring model we consider in this paper is defined by the following action\cite{Pawlowski:2013pje,Pawlowski:2013gag,Pawlowski:2014ada,Kogut:1974ag,Hasenfratz:1983ba}, 
\begin{equation}
S_0 = 
\beta  \sum_{n=1}^{L}  \big(1- \cos A_n \big)  - \sum_{n=1}^{L} \sum_{f=1}^{N_f}
\bar \chi^f_n  \left\{
 {\rm e}^{i A_n +\mu a } \, \chi^f_{n+1} -  {\rm e}^{-i A_{n-1}-\mu a } \,  \chi^f_{n-1}  
 + m  a  \, \chi^f_n \right\} , 
\end{equation}
where $\beta = {1}/{2 g^2 a}$,  $m a$, $\mu a$ 
are the inverse coupling, mass and chemical potential in the lattice unit, 
and  $L$ is the lattice size which defines the inverse temperature ($T \equiv 1/La$).
The  fermion field $\chi^f$, $\bar \chi^f$ has $N_f$ flavors and
satisfies the anti-periodic boundary conditions: 
$\chi^f_{L+1}=-\chi^f_1$, $\chi^f_0=-\chi^f_L$ 
and $\bar \chi^f_{L+1}=-\bar\chi^f_1$, $\bar\chi^f_0=-\bar\chi^f_L$. 
The auxiliary field $A_n$, 
which should couple to the vector current of the fermion $\chi^f$, $\bar \chi^f$, 
is introduced as a compact link variables ${\rm e}^{i A_n}$.
The partition function of the lattice model is  defined by the path-integration,
\begin{eqnarray}
\label{eq:path-integral-original}
Z &=& \int {\cal D} A  {\cal D}  \chi {\cal D}  \bar\chi \, \, {\rm e}^{- S_0} \nonumber\\
   &=&  \int_{-\pi}^{\pi} \prod_{n=1}^L d A_n  \, {\rm e}^{- \beta  \sum_{n=1}^{L}  \big(1- \cos A_n \big) } 
   \, { \det D [A] }^{N_f}, 
   \end{eqnarray}
where $D$ denotes the lattice Dirac operator, 
\begin{equation}
( D \chi )_n =
 {\rm e}^{i A_n +\mu a } \, \chi^f_{n+1} -  {\rm e}^{-i A_{n-1}-\mu a} \,  \chi^f_{n-1}  
 + m a \, \chi^f_n . 
 \end{equation}
The functional determinant of $D$ can be evaluated explicitly as 
\begin{equation}
\label{eq:detD-A}
 \det D \, [A] = \frac{1}{2^{L-1}} \Big[ \cosh (L  \hat \mu + i {\scriptstyle \sum_{n=1}^L } A_n) + \cosh L \hat m \Big]  
  \qquad (\hat \mu = \mu a, \, \, \hat m=\sinh^{-1} m a ).
\end{equation}
This is not real-positive in general when $\mu \not = 0$ and it has 
the property $ \left( \det D[A]\vert_{+\mu} \right)^\ast = \det D[-A]\vert_{+\mu} = \det D[A]\vert_{-\mu}$.  This fact can cause the sign problem in Monte Carlo simulations.
We consider the case of $N_f=1$ for simplicity in the following sections. 

\begin{figure}[thbp]
  \begin{center}
    \begin{tabular}{cc}
      \begin{minipage}{0.5\hsize}
        \begin{center}
          \includegraphics[clip, width=70mm]{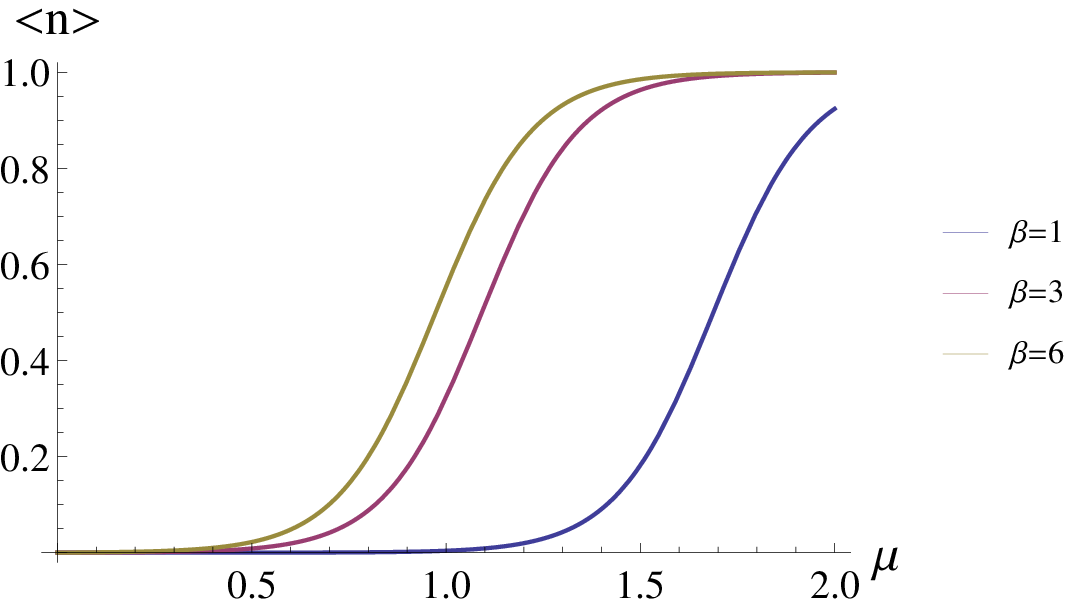}
          \hspace{1.6cm} (a) Number density
        \end{center}
      \end{minipage} 

      \begin{minipage}{0.5\hsize}
        \begin{center}
          \includegraphics[clip, width=70mm]{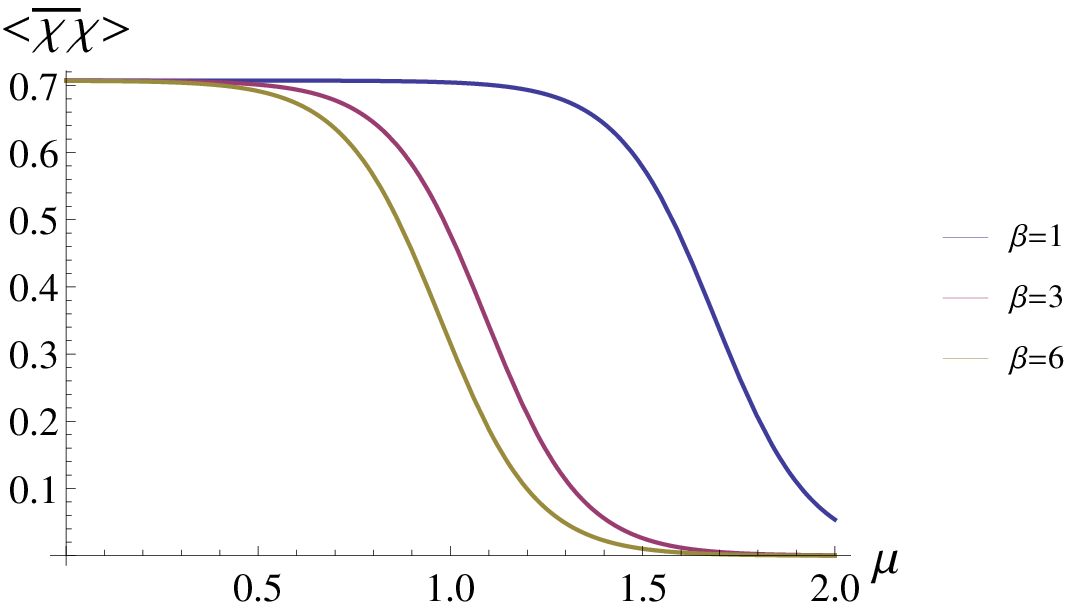}
          \hspace{1.6cm} (b) Meson condensate
        \end{center}
      \end{minipage} 
    \end{tabular}
    \caption{Exact value of the number density (a) and the scalar condensate (b) with $m=1,L=8$ at $\beta=1,3$, and $6$.}
    \label{fig:exact_n_chi_cp}
  \end{center}
\end{figure}

This lattice model is exactly solvable in the following sense. 
The path-integration over the field $A_n$ can be done 
explicitly and the exact expression of the partition function is obtained with the modified Bessel functions of the first kind as
\begin{equation}
Z= \frac{1}{2^{L-1}} \, {\rm e}^{-L\beta} \, \Big[ I_1(\beta)^L  \cosh L \hat \mu +  I_0(\beta)^L \cosh L \hat m \Big] .
\end{equation}
The number density and condensate of the fermion field are then obtained as follows:
\begin{eqnarray}
\langle n \rangle 
&\equiv& \frac{1}{La } \frac{\partial \ln Z}{\partial \mu}  \nonumber\\
&=& \frac{ I_{1}(\beta)^{L} \sinh L \hat \mu }{I_{1}(\beta)^{L} \cosh L \hat \mu + I_{0}(\beta)^{L} \cosh L \hat{m}},\\
&&\nonumber\\
\langle \bar{\chi} \chi \rangle
&\equiv&\frac{1}{La} \frac{\partial \ln Z}{\partial m} \nonumber\\
&=& \frac{ I_{0}(\beta)^{L} \sinh L \hat{m}}{ [I_{1}(\beta)^{L} \cosh L \hat\mu + I_{0}(\beta)^{L} \cosh L \hat{m} ]\cosh \hat{m}}.
\end{eqnarray}
The $\mu$-dependence of these observables are plotted in 
fig.~\ref{fig:exact_n_chi_cp} 
for $L=8$, $ma=1$, and $\beta=1,3,6$.
It shows a crossover behavior in the chemical potential $\hat \mu$ (in the lattice unit) around
$\hat \mu \simeq \hat m + \ln (I_0(\beta)/I_1(\beta))$. 
\begin{figure}[htbp]
  \begin{center} 
    \includegraphics[clip, width=80mm]{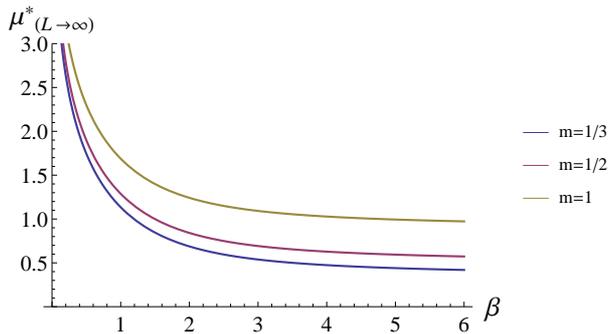} 
    \caption{$\beta$-dependence of the critical chemical potential for the large $L$ limit with $m=1/3,1/2$, and $1$.}
    \label{fig:crid_dens_largeN} 
  \end{center}
\end{figure}
In the limit $L \rightarrow \infty$, these quantities reduce to the following forms,
\be
\lim_{L \rightarrow \infty} \langle n \rangle &=& H_{1/2} \big(\hat \mu - \mu^\ast_{(L \rightarrow \infty)}\big), \label{eq:numdens_lowT}\\
\lim_{L \rightarrow \infty} \langle \bar{\chi}\chi \rangle &=& \frac{1-H_{1/2}\big(\hat \mu - \mu^\ast_{(L \rightarrow \infty)}\big)}{\cosh \hat{m}}, \label{eq:chicond_lowT}
\ee
where $H_{1/2}(x)$ is the Heaviside step function
and $\mu^\ast_{(L \rightarrow \infty)}$ is the critical density in this limit given by
$\mu^\ast_{(L \rightarrow \infty)} = \hat{m} + \ln (I_0(\beta)/I_1(\beta))$. 
fig.\ref{fig:crid_dens_largeN} shows the $\beta$-dependence of the critical density at $ma=1/3,1/2$, and $1$.


The continuum limit of the lattice model ($a \rightarrow 0$)  may be defined  at a finite temperature
as the limit: $\beta = {1}/{2 g^2 a} \rightarrow \infty$, $L = {1}/{Ta}  \rightarrow \infty$, while 
$\beta / L = T/2 g^2$ fixed. 
%
In this limit, the partition function scales as 
\begin{equation}
Z \longrightarrow \frac{1}{2^{L-1}}\left(\frac{1}{2\pi\beta}\right)^{L/2} 
{\rm e}^{\frac{3 g^2}{4T}}  
\left( \cosh \frac{\mu}{T}  + {\rm e}^{\frac{g^2}{T}}  \cosh \frac{m}{T} \right) , 
\end{equation}
and the continuum limits of $\langle n \rangle$ and $\langle \bar \chi \chi \rangle$
are obtained as follows:
\begin{eqnarray}
%
\lim_{a \rightarrow 0} \, \langle n \rangle 
&=& \frac{\sinh \frac{\mu}{T} }
{ \cosh \frac{\mu}{T}+ {\rm e}^{\frac{g^2}{T}} \cosh \frac{m}{T} } , 
\nonumber\\
\lim_{a \rightarrow 0} \, \langle \bar \chi \chi \rangle 
&=& \frac{ {\rm e}^{\frac{g^2}{T}} \sinh \frac{m}{T}}{ \cosh \frac{\mu}{T} + {\rm e}^{\frac{g^2}{T}}  \cosh \frac{m}{T} } . 
\end{eqnarray}
From these results, 
one can see that the model shows a crossover behavior in the chemical potential $\mu$ 
for a non-zero temperature $T > 0$, while  in the zero temperature limit $T=0$, 
it shows a first-order transition at the critical chemical potential $\mu_c   = m + g^2$.  
We note that at the zero temperature $T=0$, 
the number density $\langle n \rangle$ vanishes identically 
for $\mu \le \mu_c$, which is sometimes called as the Silver-Blaze behavior \cite{Cohen:2003kd}.


%
%



\subsection{Thirring model complexified on Lefschetz thimbles}
\label{sec:appliation_LTM}

Next we consider the complexification of  the above lattice model and reformulate 
the defining path-integral of eq.~(\ref{eq:path-integral-original}) 
by the complex integrations over Lefschetz thimbles.
In the complexification,  
%
the field variables $A_n $ are extended to complex variables $z_n \, (\in \mathbb{C}^L)$ and 
the action is extended to the complex function given by 
$S[z] = \beta \sum_{n=1}^L (1- \cos z_n)  - \ln \det D [z]$.
Then, for each critical point $z (=\{ z_n \}) = \sigma$ given by the stationary condition,
 \begin{equation}
 \label{eq:stationary-condition}
 \beta \sin z_n - i 
\frac{\sinh (L  \hat \mu + i {\scriptstyle \sum_{\ell=1}^L } z_\ell) }
{\cosh (L  \hat \mu + i {\scriptstyle \sum_{\ell=1}^L } z_\ell) + \cosh L \hat m }
 =0
 \qquad (n=1,\cdots,L), 
  \end{equation}
the thimble $\mathcal J_\sigma$ 
is defined as the union of all the (downward) flows given by the solutions of the gradient flow equation 
\begin{eqnarray}
\label{eq:downward-flow-equation}
\frac{d}{dt} z_n(t) =  \frac{ \partial \bar S [ \bar z] }{\partial \bar z_n} 
\quad  
( t \in \mathbb{R} )  
\qquad \text{s.t.} \quad
z(-\infty) =\sigma.
\end{eqnarray}
The thimble so defined is an $L$-dimensional real submanifold in $\mathbb{C}^L$.
Then, according to Picard-Lefschetz theory (complexified Morse theory),  the original path-integration region 
${\mathcal C}_\mathbb{R} \equiv [ -\pi , \pi]^L$
can be replaced with a set of 
Lefschetz thimbles\footnote{Here we assume  
${\mathcal C}_\mathbb{R} \equiv ([ - \pi+i \infty, -\pi] \oplus  [ -\pi , \pi] \oplus [\pi, \pi+i\infty])^L $.},
\begin{equation}
{\mathcal C}_\mathbb{R}
= \sum_{\sigma}  n_\sigma {\cal J}_\sigma , 
\end{equation}
where 
$n_\sigma$ stands for the intersection number between 
${\mathcal C}_\mathbb{R}$ 
and the other $L$-dimensional real submanifold ${\cal K}_\sigma$ 
of $\mathbb{C}^L$ associated to the same critical point $\sigma$, 
defined as the union of all the gradient flows  s.t. $z(+\infty) = \sigma$.
Namely, the partition function and the correlation functions of the lattice model can 
be expressed by the formulae,
\begin{eqnarray}
\label{eq:partion-function-by-thimbles}
Z &=& \sum_{\sigma \in \Sigma}  n_{\sigma}  \, {\rm e}^{- S[\sigma]}  
\, Z_{\sigma}, 
\, \qquad \qquad
Z_{\sigma} \equiv
\int_{{\cal J}_\sigma} {\cal D}[z] \, 
{\rm e}^{-(S[z] -S[\sigma])},  \\
\langle O[z] \rangle &=& 
\frac{1}{Z} \sum_{\sigma \in \Sigma}  
n_{\sigma}  \, {\rm e}^{- S[\sigma]}  \, Z_{\sigma} \, \langle O[z] \rangle_{\sigma} , 
\quad
\langle O[z] \rangle_{\sigma}
\equiv \frac{1}{\, Z_{\sigma}} \, 
\int_{{\cal J}_\sigma} {\cal D}[z] \, {\rm e}^{-(S[z] -S[\sigma])} O[z] .
\nonumber\\
\end{eqnarray}

\begin{figure}[thbp]
  \begin{center}
    \begin{tabular}{cc}
      \begin{minipage}{0.5\hsize}
        \begin{center}
          \includegraphics[clip, width=70mm]{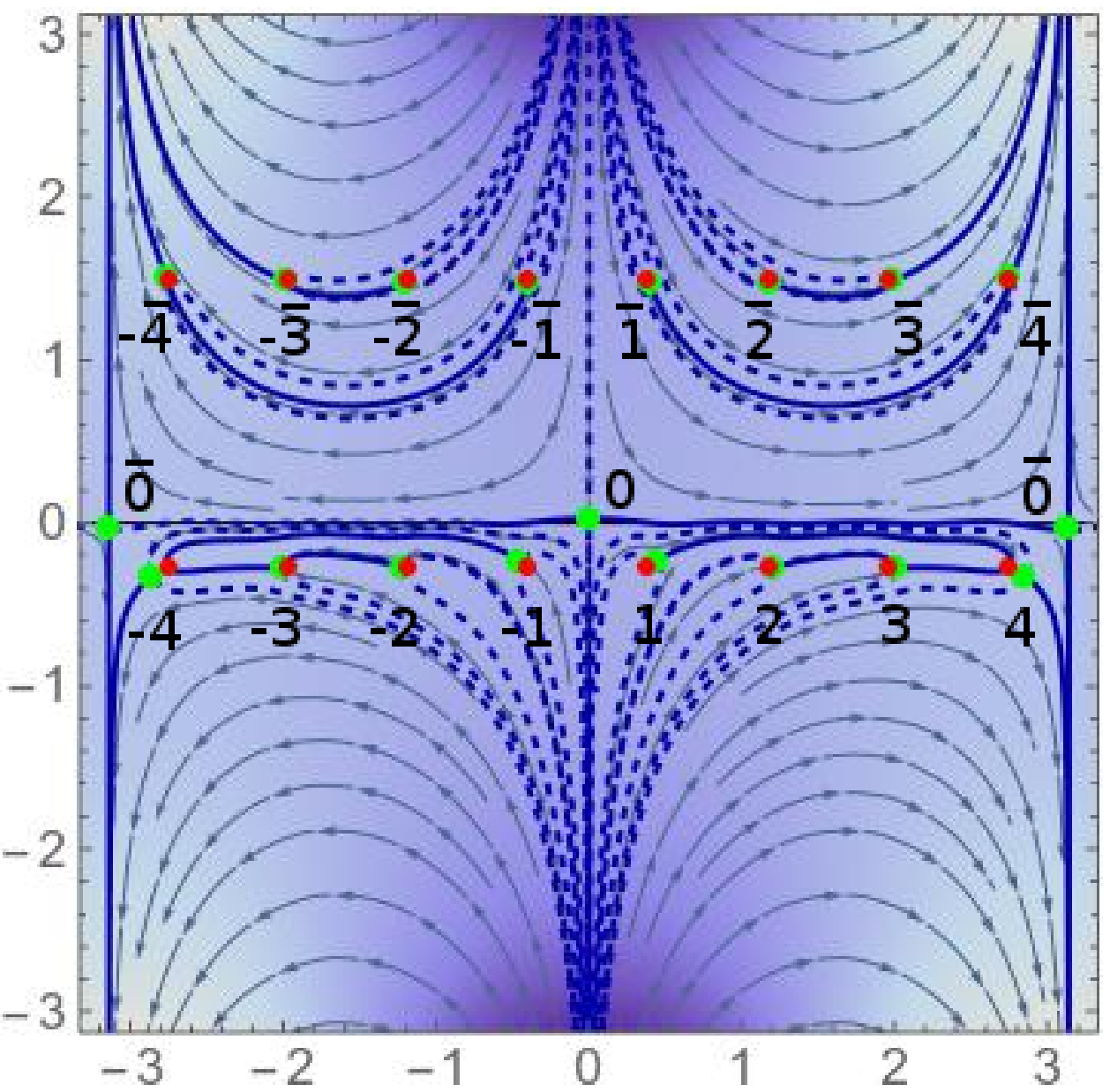}
          \hspace{1.6cm} (a) $\mu=0.6$
        \end{center}
      \end{minipage} 

      \begin{minipage}{0.5\hsize}
        \begin{center}
          \includegraphics[clip, width=70mm]{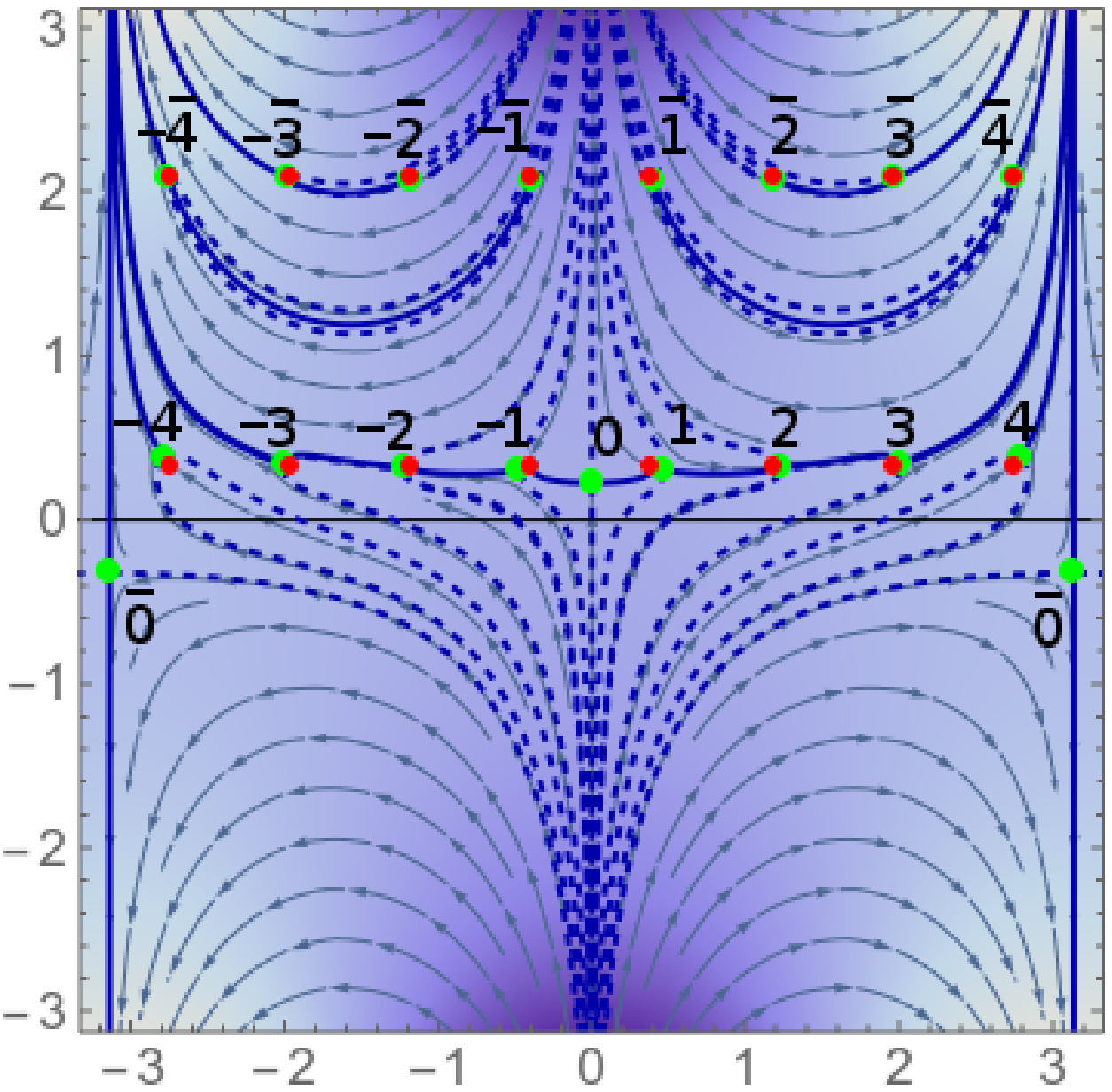}
          \hspace{1.6cm} (b) $\mu=1.2$
        \end{center}
      \end{minipage} 
    \end{tabular}
    \caption{
    The critical points given by the solutions of eq.(\ref{eq:critical-points-dominant})
    for $L=8$, $\beta=3$, $m a = 1$ and $\mu a =0.6,1.2$.
    The critical points (green points), the thimbles(blue lines: downward flows, 
    blue dotted lines: upward flows) and the zeros of $\det D[z]$(red points)
    are shown 
    in the complex plane $z \in \mathbb{C}$ (which parametrizes 
    the field subspace of $z_n=z  \, (n=1, \cdots, L)$ in $\mathbb{C}^L$).
    The numbers in the figure are used to label the critical points. The most dominant thimble
    is ${\cal J}_{\sigma_0}$, whose value of the action $S[\sigma]$ is closest to that of the classical vacuum.
    }
    \label{fig:conf_sp}
  \end{center}
\end{figure}

It is not straightforward in general to find all the critical points $\{\sigma\}$
and to work out the intersection numbers $\{ n_\sigma\}$ of the associated Lefschetz thimbles 
$\{{\cal J}_\sigma \}$. Fortunately, in our lattice model, we can obtain 
all the solutions of the stationary condition eq.~(\ref{eq:stationary-condition})
and therefore all the critical points. In the separated paper\cite{Fujii:2015bua}, 
we have shown that 
the critical points can be classified by an integer $n_- (=0, 1, \cdots, L/2-1)$ as 
\begin{eqnarray}
z_n &=&  \left\{ \begin{array}{l} z \\ \pi-z \end{array} \right.  
\qquad (n=1, \cdots, L),  \\
0 &=&  \beta \sin z  - \frac{i \sinh [ L \hat \mu + i (L-2 n_-) z  ]}
{\cosh[L \hat \mu + i (L-2 n_-) z]+(-1)^{n_-} \cosh(L\hat m)}, 
\end{eqnarray}
where $n_-$ is defined as the number of the components $z_n$ 
which take the value $\pi - z$. 
Moreover, 
by inspecting the solutions of the gradient flow equation, 
the values of the action at the critical points $\{ S[\sigma] \}$  and the Stokes phenomena, 
we have identified the set of the thimbles which contribute to the path-integral 
for given parameters, $L$, $\beta$,  $m$ and $\mu$. 
%
Especially, we found that the dominant thimbles are associated 
with the critical points of the type $n_-=0$, 
\begin{eqnarray}
\label{eq:critical-points-dominant}
&& z_n = z  \quad (n=1, \cdots, L),    \nonumber\\
&&  
\beta \sin z - \frac{i \sinh(L (\hat \mu + i z) )}{\cosh(L (\hat \mu + i z))+\cosh(L\hat m)} = 0.
\end{eqnarray}
These critical points are shown in fig.\ref{fig:conf_sp} for $\beta=3$, $ma=1$, $L=8$ and $\mu a =0.6$ 
in the complex plane of $z \in \mathbb{C}$ which parameterizes
the field subspace of $ z_n = z \, (n=1, \cdots, L)$ in $\mathbb{C}^L$.  
We denote these critical points by the labels  $\sigma_i$ and $\bar\sigma_i$ with 
$i=0,\pm 1, \cdots, \pm L/2$ also shown in fig.~\ref{fig:conf_sp} (for the case $L=8$).

We note that some of the thimbles terminate at the zeros of the fermion determinant, 
\begin{equation}
\left. \det D[z] \right\vert_{z=z_{\rm zero}} = 0, \qquad
z_{\rm zero}= i (\hat \mu \pm \hat{m} ) + \frac{2 n +1}{L}\pi \quad ( n \in {\mathbb Z} \,\, \text{mod} \,\, L), 
\end{equation}
which are also shown in the figure.  

Among these thimbles associated with the critical points given by eq.~(\ref{eq:critical-points-dominant}), the most dominant thimble is the thimble ${\cal J}_{\sigma_0}$,  
which is labeled by $0$ in the figure. 
It turns out that its value of the action $S[\sigma_0]$ is closest to that of the classical vacuum of the model. 
In the following numerical study, we consider  this most dominant thimble ${\cal J}_{\sigma_0}$.


\section{Hybrid Monte Carlo study of the Thirring model on the thimble ${\cal J}_{\sigma_0}$}

In this section, we describe our numerical simulations
of the Thirring model performed on the single thimble ${\cal J}_{\sigma_0}$. 
First, we review the Lefschetz thimble HMC method proposed in ref.\cite{Fujii:2013sra}, 
and discuss a few improvements of the method necessary in applying to the (fermionic) 
Thirring  model.  Secondly, we summarize the simulation parameter details. 
Lastly, we present and discuss our simulation results.

%
%
%

\subsection{Simulation method : Hybrid Monte Carlo on Lefschetz thimbles} 
\label{sec:Lefschetz_HMC}

The hybrid Monte Carlo (HMC) algorithm on Lefschetz thimbles proposed in \cite{Fujii:2013sra}
is a Monte Carlo method to evaluate the path-integral of an observable $O[x]$ 
over a given thimble 
${\cal J}_\sigma$, 
\begin{equation}
\langle O \rangle_{\sigma}
\equiv \frac{1}{\, Z_{\sigma}} \, 
\int_{{\cal J}_\sigma} {\cal D}[z] \, {\rm e}^{-(S[z] -S[\sigma])} O[z], 
\end{equation}
where the functional measure ${\cal D}[z]$ along the thimble ${\cal J}_\sigma$ 
is specified as $ \left. dz^L \right\vert_{{\cal J}_\sigma} = d (\delta \xi)^L \det U_z$
by the orthonormal basis of tangent vectors $\{ U_z^\alpha  \vert (\alpha=1,\cdots, L) \}$ 
which span the tangent space as $ \delta z = U_z^\alpha \delta \xi^\alpha $ $( \delta z \in \mathbb{C}^L, \delta \xi \in \mathbb{R}^L )$.
In this HMC algorithm, 
a series of  field configurations $\{ z^{(k)} \} \, (k=1,\cdots, N_{\rm conf} )$ are generated with the real-positive weight 
$ \left. {\rm e}^{-(S[z] -S[\sigma])} \right\vert_{{\cal J}_\sigma}$
through the Molecular dynamics steps constrained to the thimble and the Metropolis accept/reject procedure, 
while the residual complex phase factor 
${\rm e}^{i \phi_z} = \det U_z$ is reweighed to the observable as 
\begin{equation}
\langle O \rangle_{{\sigma}} = \lim_{N_{\rm conf} \rightarrow \infty} \frac{\langle 
e^{ i \phi_{z}} O  \rangle'}{\langle e^{ i \phi_{z}} \rangle'}  \, ; \qquad
\langle X \rangle'  = 
\frac{1}{N_{\rm conf}} \sum_{k=1}^{N_{\rm conf}} X[z^{(k)}] . 
\end{equation}
In the algorithm, 
any field configuration $z$ on the thimble and the associated tangent vectors 
$\{ V_{z \, n}^\alpha \} (\alpha=1,\cdots, L)$ are computed by solving 
the flow equations\footnote{In the following, we will use the abbreviation
$\partial / \partial z_n = \partial_n$, $\partial / \partial \bar z_n = \bar  \partial_n$.}
\begin{equation}
\label{eq:gradient-flow-equations}
\frac{d  }{dt} z_{n}(t) = 
\bar \partial_n \bar S[\bar z], \qquad
\frac{d  }{dt} V^{\alpha}_{z\,n}(t) = 
\bar \partial_n \bar \partial_m \bar S[\bar z] \, \bar{V}^{\alpha}_{z\, m}(t), 
\end{equation}
assuming
that the solutions take the asymptotic forms in the sufficient past at $t=t_0$ ($t_0 <0, \vert t_0 \vert \gg 1$) as
\begin{equation}
z_{n}(t_{0}) = z_{\sigma n} + 
v^{\alpha}_{n} \exp(\kappa^{\alpha} t_{0}) e^{\alpha}, 
\qquad
V^{\alpha}_{z \, n}(t_{0}) = 
v^{\alpha}_{n} \exp (\kappa^{\alpha} t_{0}). 
\end{equation}
Here $e^\alpha$ ($\alpha=1,\cdots,L$) is a real vector 
($e^{\alpha} \in {\mathbb R}$; ${\scriptstyle \sum_{\alpha=1}^L}e^{\alpha}e^{\alpha}=L$), 
and
$v^{\alpha}_{n}$  ($\alpha=1,\cdots,L$) are the orthonormal tangent vectors 
at the critical point $\sigma$ which
factorize the Hesse matrix ${\cal K}_{nm} \equiv \partial_n \partial_m S[z_{\sigma}]$
with the real-positive diagonal elements $\kappa^\alpha$ ($\alpha=1,\cdots,L$): 
$v^{\alpha}_{n} {\cal K}_{nm}v^{\beta}_{m}=\kappa^{\alpha}\delta^{\alpha \beta} $.
By this procedure, one can parameterize any field configuration $z$ on the thimble
by the set of the parameters,  the flow-direction vector 
$e^{\alpha}$ 
and the flow-time $t' = t-t_0$, defining 
a map $(e^\alpha, t') \rightarrow z \in {\cal J}_\sigma$ as
\begin{equation}
z_n[e, t'] = z_n(t) \vert_{t=t'+t_0}.
\end{equation}
We employ the 4th-order Runge-Kutta method to solve the flow equations
and use \texttt{Diag} package\cite{Hahna:2006} 
to perform the factorization of the Hesse matrix.
The molecular dynamics is then formulated as a constraint dynamical system 
and solved by the constraint-preserving second-order symmetric integrator as 
\begin{eqnarray}
w^{i+\frac{1}{2}} 
&=& w^{i} - (1/2) \Delta \tau \,  \bar \partial  \bar S[\bar z^{i}] 
                 - (1/2) \Delta \tau \, i  V^\alpha[z^{i},\bar z^{i}] \lambda^a_{[r]} ,  \\
                 z^{i+1} &=& z^{i} + \Delta \tau \, w^{i+\frac{1}{2}} , \\
w^{i+1} 
&=& w^{i+\frac{1}{2} } - (1/2) \Delta \tau \, \bar \partial  \bar S[\bar z^{i+1}] 
                 - (1/2) \Delta \tau \, i  V^\alpha[z^{i+1},\bar z^{i+1}] \lambda^a_{[v]} ,  
\end{eqnarray}                 
where  $\lambda^a_{[r]}$ and $\lambda^a_{[v]}$ are fixed by imposing 
the constraints, 
\begin{eqnarray}
z^{i+1} &=& z[e^{(i+1)},t^{' (i+1)}],  \nonumber \\
\label{eq:constraints-in-molecular-dynamics}
w^{i+1} &=& V^\alpha[z^{i+1},\bar z^{i+1}] w^{\alpha (i+1)}, \quad  
w^{\alpha (n+1)} \in \mathbb{R},                              
\end{eqnarray}
respectively.
The Metropolis accept/reject procedure is performed
with the conserved Hamiltonian
$ H=\frac{1}{2} \bar w_n w_n + \frac{1}{2} \left\{S[z]+\bar S[\bar z] \right\}$.

In the lattice Thirring model we are considering, 
the given thimble ${\cal J}_\sigma$ can terminate at the zeros of fermion determinant.
In such a case, 
the flow reaches a zero within finite time and the flow time $t' (=t-t_0)$ is bounded. 
Moreover, the force terms in the flow equations become quite large in the vicinity of the zero.
These points cause problems in solving the flow equations or in solving the Molecular dynamics
with finite time steps. To improve this situation and to achieve the necessary precision of the solutions, we implement in this work the adaptive step size in the 4th-order Runge-Kutta method: 
we simply adjust the step size $\Delta t$ depending on the size of the force terms 
$F_{n}[z]$ as
$\left| F_{n}[z] \right| \cdot \Delta t = L \cdot {\rm const.}.$
In this respect, an estimate of the error of the solutions can be obtained by using 
$R= \vert
        \partial \bar{S}/ \partial \bar{z}_{n} - V_{z \, n}^\alpha \kappa^\alpha e^\alpha 
        \vert^2 /2L$, 
which should vanish for an exact solution.
%
We also introduce and adjust 
a scale parameter $\lambda$ as $z_{n} \rightarrow \lambda z_{n}$
to keep the values of the diagonal elements $\kappa^\alpha$ of the Hesse matrix
in a reasonable range,  
for otherwise 
the exponential growth of the field configurations could be very rapid
with a finite step size, 
the errors in the solutions of the flow equations 
eqs.~(\ref{eq:gradient-flow-equations}) could
become out of control, 
and 
the iterate method to solve the constraints 
eqs.~(\ref{eq:constraints-in-molecular-dynamics}) could not converge.

\subsection{Simulation details} \label{sec:sim_details}
The parameter sets in our simulations are summerized 
as follows. 
The base simulations were performed for $ma =1$, $\beta=1,3,6$ on the lattice $L=4,8$
in order to measure and examine the averages of the residual phase, 
number density and scalar condensate. 
A series of simulations for $L=8,16, 32$ with $L (m a)  =16$ and $\beta (m a) = 2, 3$ 
were done in the study of the continuum limit behavior, 
and 
a series of simulations for $L=4, 8,12,16, 24, 32$ with $\beta=3$, $m a =1$ were used for the study of the low-temperature limit behavior.
For each parameter sets, the chemical potential was varied in the range $\mu a \in [0.0, 2.0]$
with the increment $0.2$.

In solving the flow equations by the Runge-Kutta method,  we set $t_{0}=-4$.
The initial values of 
the number of steps and the step size are $N_{t}=20$ and $\Delta t =0.1$, respectively.
The scale parameter $\lambda$ is chosen in the range $0.05 \le \lambda \le 0.1$.
With these parameters, the condition $R  < 10^{-5}$ was satisfied.

For the Molecular dynamics, the trajectory length and  the number of steps are set to $\tau = 0.5$ and $N_{\tau}=10$, respectively. 
We generated 1,000 configurations for all the parameter sets and estimated errors using the jackknife method with a bin per 20 configurations.



\subsection{Simulation results}

\begin{figure}[htbp]
  \begin{center}
    \begin{tabular}{cc}
      \begin{minipage}{0.5\hsize}
        \begin{center}
          \includegraphics[width=80mm,angle=0]{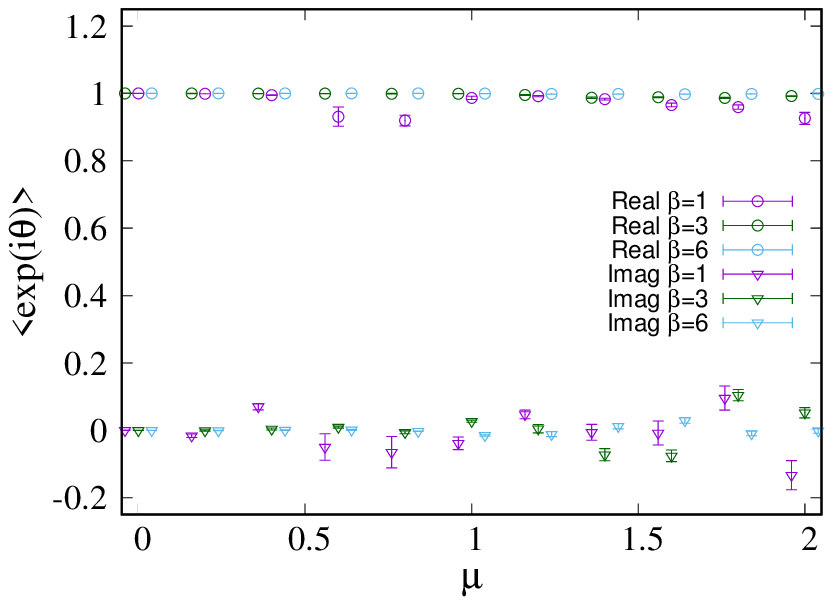}
          \hspace{1.6cm} (a) $L=4$
        \end{center}
      \end{minipage} 
      \begin{minipage}{0.5\hsize}
        \begin{center}
          \includegraphics[width=80mm,angle=0]{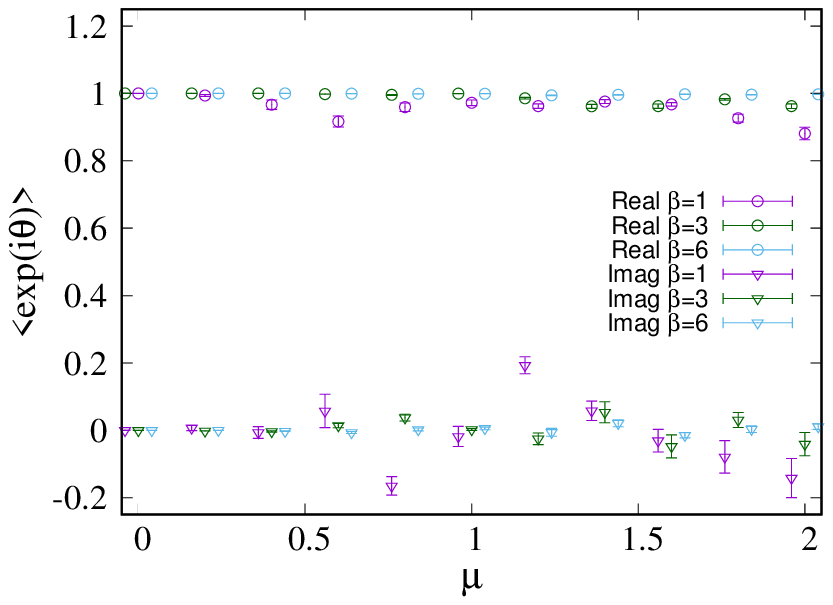}
          \hspace{1.6cm} (b) $L=8$
        \end{center} 
      \end{minipage} 
    \end{tabular}
    \caption{The averages of the residual phase factor for $ma =1$, $\beta=1,3,6$ and $L=4, 8$.}
    \label{fig:phasem10b3_bdep}
  \end{center}
\end{figure}

First of all, we show in fig.~\ref{fig:phasem10b3_bdep}
the result on the averages of the residual phase for  
$ma=1$, $\beta=1,3,6$ and $L=4,8$.
The average ${\rm Re} \langle \exp(i \theta) \rangle$ 
sometimes deviates from unity, but it  stays greater than $0.8$ almost always. 
The similar results were observed for the larger lattice sizes $L=12, 16, 24, 32$.
From these results,  we can say that the reweighting should work for this model 
with our choice of the parameter sets.

\begin{figure}[htbp]
  \begin{center}
    \begin{tabular}{cc}
      \begin{minipage}{0.5\hsize}
        \begin{center}
          \includegraphics[width=80mm,angle=0]{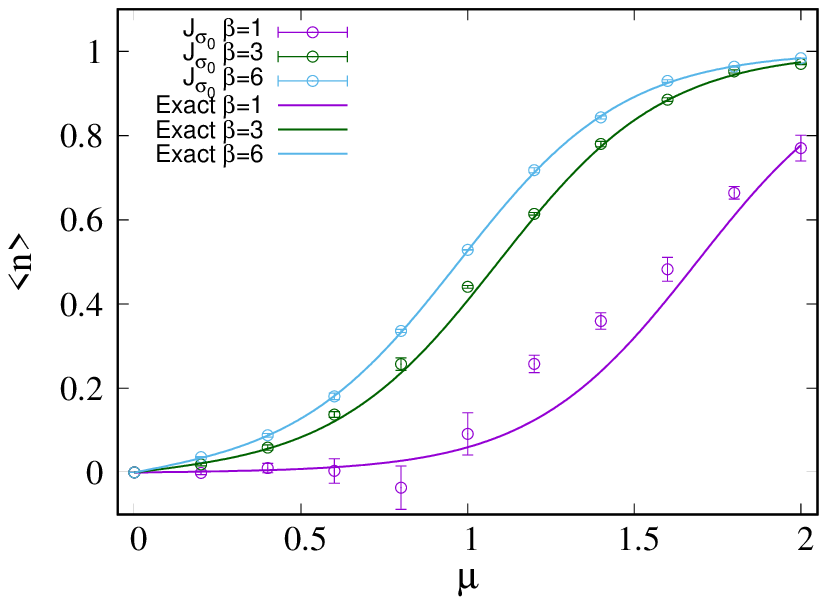}
          \hspace{1.6cm} (a) Number density
        \end{center}
      \end{minipage} 
      \begin{minipage}{0.5\hsize}
        \begin{center}
          \includegraphics[width=80mm,angle=0]{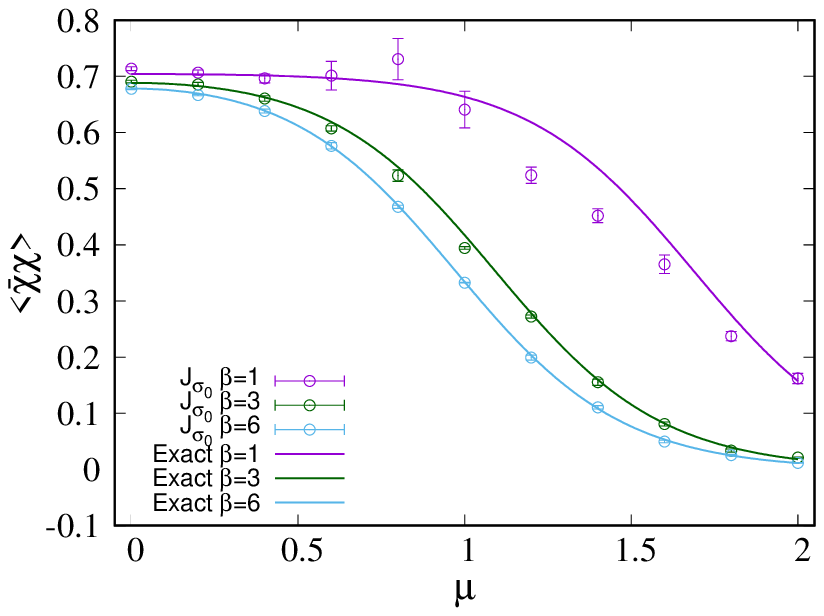}
          \hspace{1.6cm} (b) Scalar condensate
        \end{center} 
      \end{minipage} 
    \end{tabular}
    \caption{The number density and scalar condensate at $ma =1$ and $\beta=1,3,6$ on the lattice $L=4$.}
    \label{fig:obsL4m10b3_bdep}
  \end{center}
\end{figure}

\begin{figure}[htbp]
  \begin{center}
    \begin{tabular}{cc}
      \begin{minipage}{0.5\hsize}
        \begin{center}
          \includegraphics[width=80mm,angle=0]{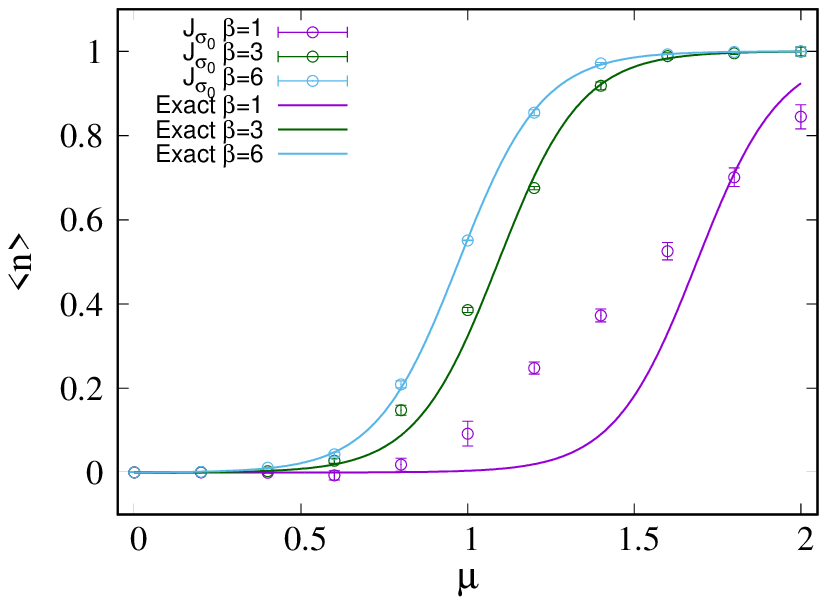}
          \hspace{1.6cm} (a) Number density
        \end{center}
      \end{minipage} 
      \begin{minipage}{0.5\hsize}
        \begin{center}
          \includegraphics[width=80mm,angle=0]{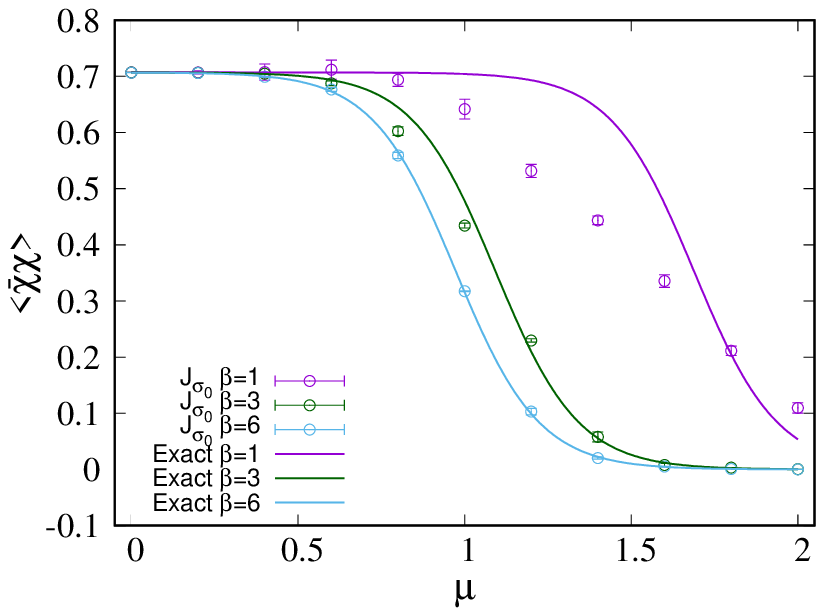}
          \hspace{1.6cm} (b) scalar condensate
        \end{center} 
      \end{minipage} 
    \end{tabular}
    \caption{The number density and scalar condensate at $ma=1$ and $\beta=1,3,6$ on the lattice $L=8$.}
    \label{fig:obsL8m10b3_bdep}
  \end{center}
\end{figure}

We next show the results of the number density and the scalar condensate 
for $L=4$ in fig. \ref{fig:obsL4m10b3_bdep} and 
for $L=8$ in fig. \ref{fig:obsL8m10b3_bdep}, respectively.
At the larger inverse couplings $\beta=3,6$,  our numerical results are in good agreement with the exact results. But at the smaller inverse coupling $\beta=1$,  discrepancies are observed in the crossover region on the both lattice sizes $L=4, 8$.
According to the analysis in \cite{Fujii:2015bua}, especially the plots in fig.~9, 
the subdominant thimbles ${\cal J}_{\sigma_1}$, ${\cal J}_{\bar \sigma_1}$ should contribute to 
the observables in the ranges of 
$[0.55, 2.1]$, $[0.7, 1.5]$, $[0.8, 1.2]$ for $\beta=1, 3, 6$ with $L=4$, respectively. 
The discrepancies observed at $\beta=1$ for $L=4~(8)$ in our simulations 
clearly indicate that this is indeed the case 
and ${\cal J}_{\sigma_1}$, ${\cal J}_{\bar \sigma_1}$ have  substantial contributions.
These results are also quite consistent with the analysis of the single-thimble approximation
shown in fig.~10 of \cite{Fujii:2015bua} using the ``uniform-field model'' .

\begin{figure}[htbp]
  \begin{center}
    \begin{tabular}{cc}
      \begin{minipage}{0.5\hsize}
        \begin{center}
          \includegraphics[width=80mm,angle=0]{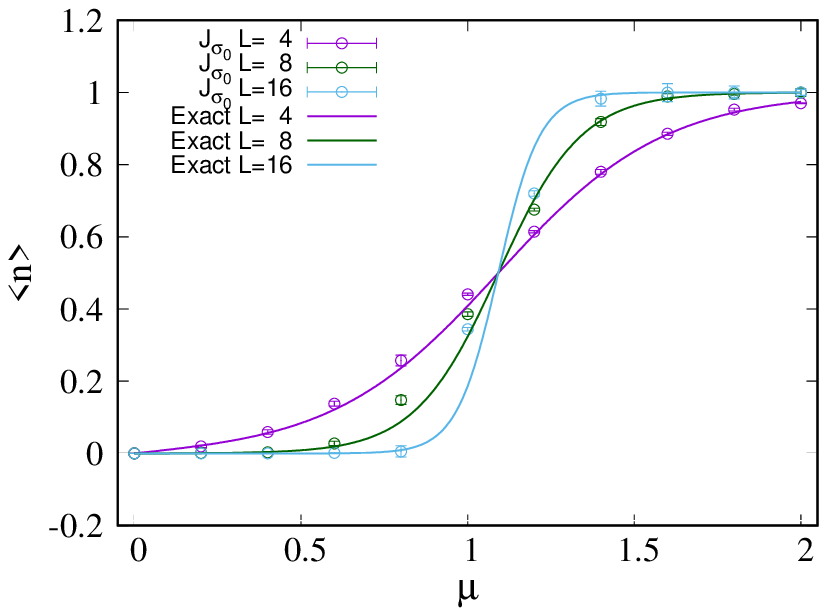}
          \hspace{1.6cm} (a) Number density
        \end{center}
      \end{minipage} 
      \begin{minipage}{0.5\hsize}
        \begin{center}
          \includegraphics[width=80mm,angle=0]{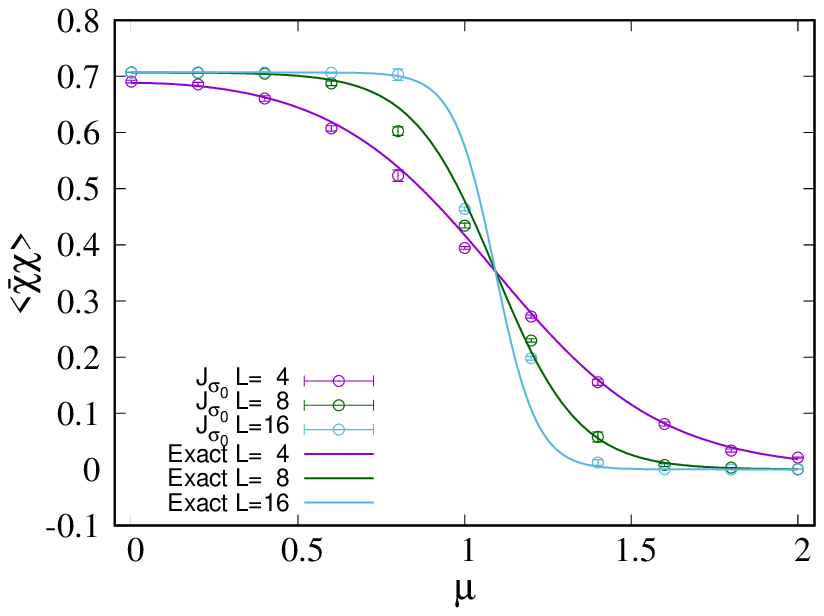}
          \hspace{1.6cm} (b) scalar condensate
        \end{center} 
      \end{minipage} 
    \end{tabular}
    \caption{$L$-dependence of the number density and scalar condensate at $ma=1$, $\beta = 3$.}
    \label{fig:obsm10b3_lT}
  \end{center}
\end{figure}

\begin{figure}[htbp]
  \begin{center}
    \begin{tabular}{cc}
      \begin{minipage}{0.5\hsize}
        \begin{center}
          \includegraphics[width=80mm,angle=0]{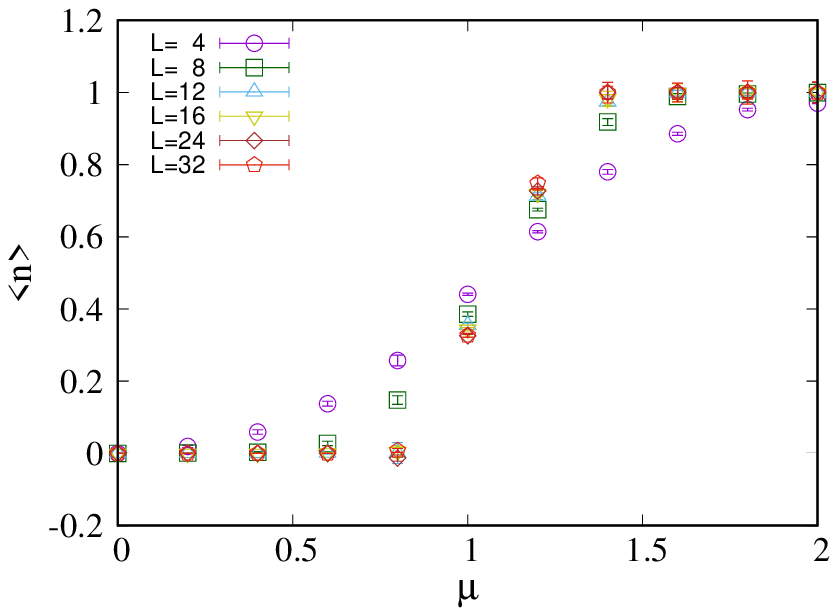}
          \hspace{1.6cm} (a) Number density
        \end{center}
      \end{minipage} 
      \begin{minipage}{0.5\hsize}
        \begin{center}
          \includegraphics[width=80mm,angle=0]{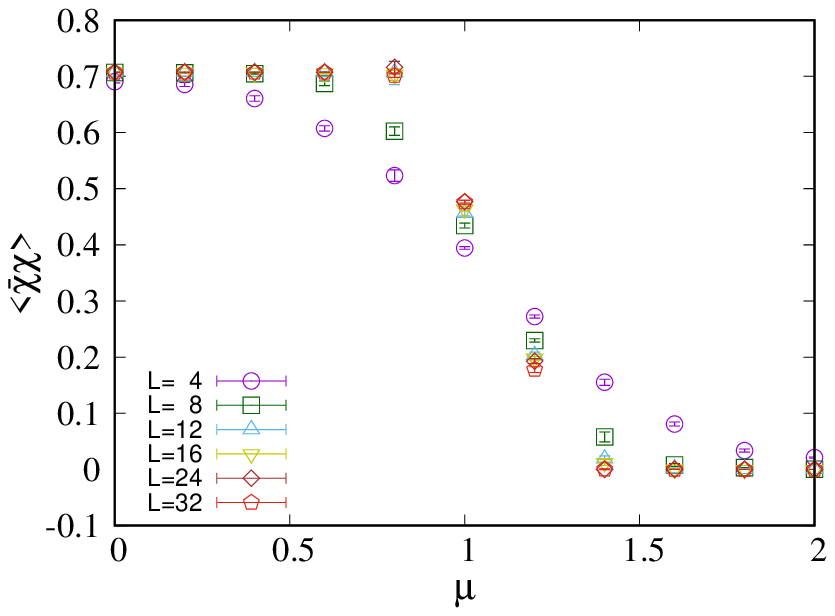}
          \hspace{1.6cm} (b) scalar condensate
        \end{center} 
      \end{minipage} 
    \end{tabular}
    \caption{Low temperature limit of the number density and scalar condensate at $ma=1,\beta=3$.}
    \label{fig:obsm10b3_Ldep}
  \end{center}
\end{figure}

In fig. \ref{fig:obsm10b3_lT}, on the other hand, 
we show the lattice size dependence of the number density and scalar condensate at $ma =1$ and $\beta=3$.  
We find that the agreement  between the numerical and exact results
gets worse as $L$ increases from $L=4$. 
The discrepancies become significant for the larger lattice sizes, $L=16, 24, 32$, 
while the contributions of the thimble ${\cal J}_{\sigma_0}$ seem saturated 
at about $L=12$ as shown in fig.~\ref{fig:obsm10b3_Ldep}.
These results on the lattice size dependence 
are quite consistent with the analysis shown in fig.~12 of \cite{Fujii:2015bua} 
based on the ``uniform-field model''.

%
%
%
%

\begin{figure}[htbp]
  \begin{center}
    \begin{tabular}{cc}
      \begin{minipage}{0.5\hsize}
        \begin{center}
          \includegraphics[width=80mm,angle=0]{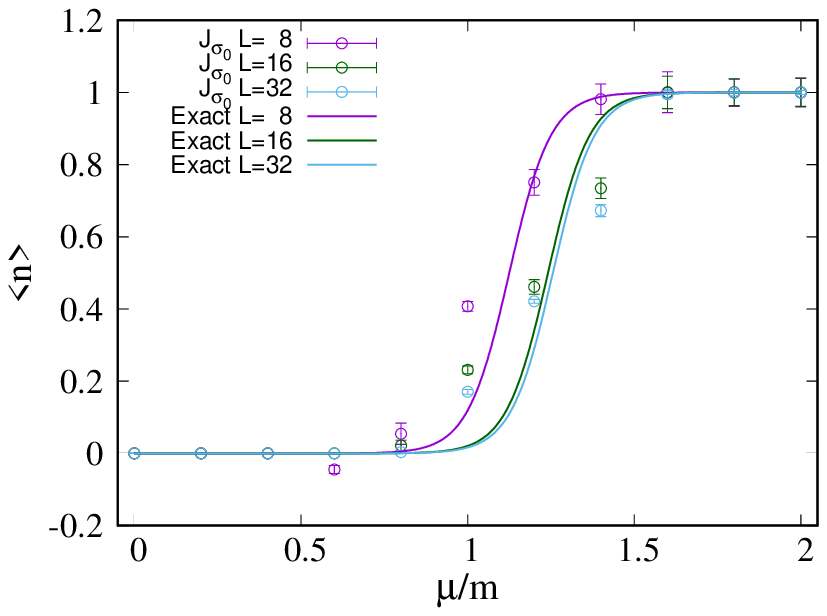}
          \hspace{1.6cm} (a) Number density
        \end{center}
      \end{minipage} 
      \begin{minipage}{0.5\hsize}
        \begin{center}
          \includegraphics[width=80mm,angle=0]{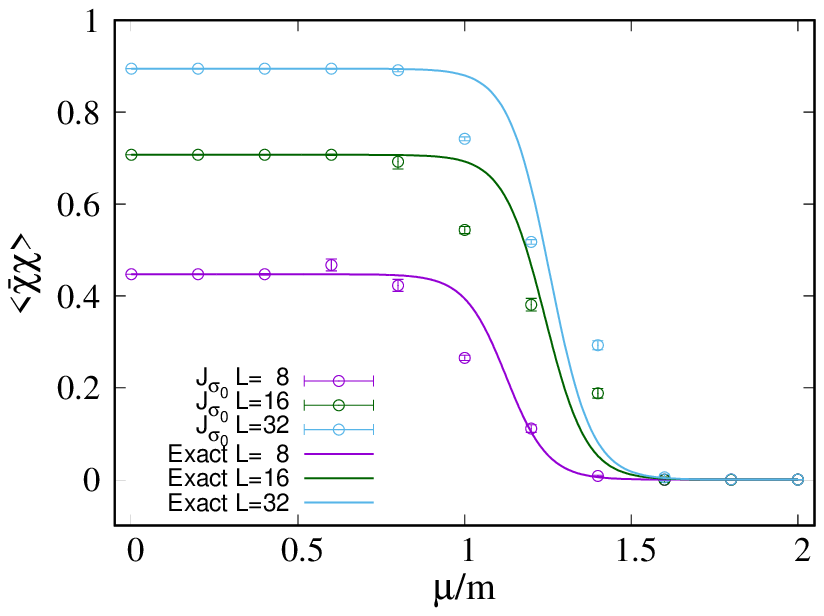}
          \hspace{1.6cm} (b) scalar condensate
        \end{center} 
      \end{minipage} 
    \end{tabular}
    \caption{Continuum limit of the number density and scalar condensate at $L m=16$ and $\beta m=2$.
We simulated with  $8,16$ and $32$ lattice sites.}
    \label{fig:obsL16m10b2_cont}
  \end{center}
\end{figure}

\begin{figure}[htbp]
  \begin{center}
    \begin{tabular}{cc}
      \begin{minipage}{0.5\hsize}
        \begin{center}
          \includegraphics[width=80mm,angle=0]{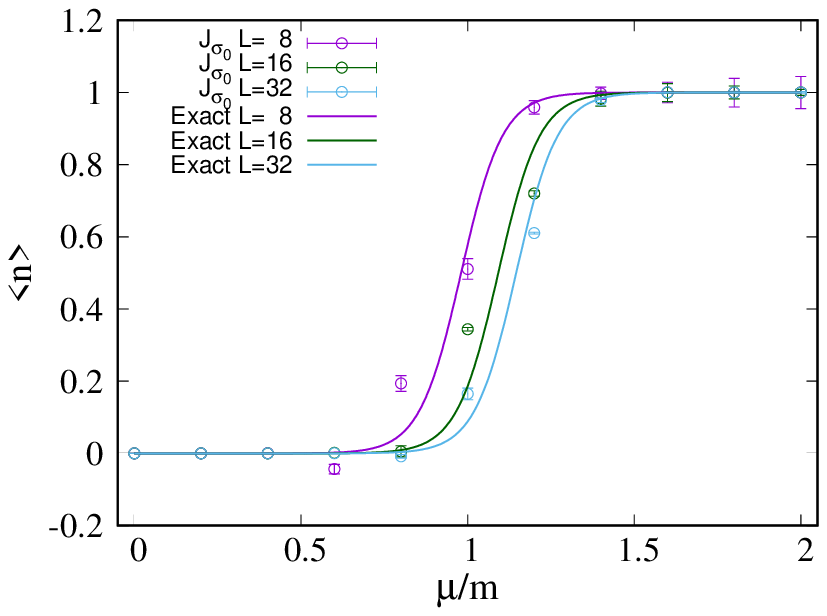}
          \hspace{1.6cm} (a) Number density
        \end{center}
      \end{minipage} 
      \begin{minipage}{0.5\hsize}
        \begin{center}
          \includegraphics[width=80mm,angle=0]{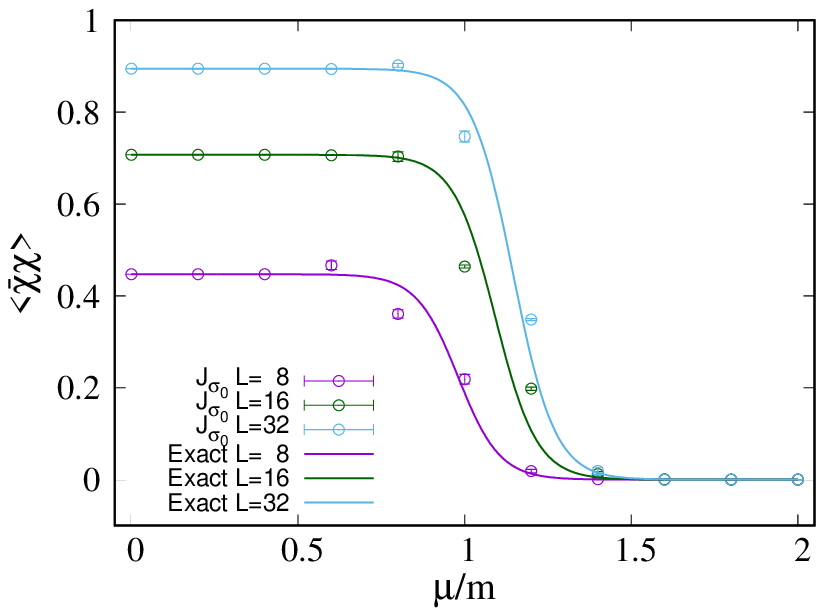}
          \hspace{1.6cm} (b) scalar condensate
        \end{center} 
      \end{minipage} 
    \end{tabular}
    \caption{Continuum limit of the number density and scalar condensate at $L m=16$ and $\beta m=3$.
We simulated with  $8,16$ and $32$ lattice sites.}
    \label{fig:obsL16m10b3_cont}
  \end{center}
\end{figure}

Finally, in figs.~\ref{fig:obsL16m10b2_cont} and \ref{fig:obsL16m10b3_cont}, we show
the results on the continuum limit at a fixed temperature. 
We find that the discrepancies observed in the crossover 
region persist in this limit. It seems that the size of the discrepancy scales, too.

%

\section{Summary and discussion}


In this paper, we have applied the Lefschetz thimble method 
to the one-dimensional lattice Thirring model at finite density 
and performed HMC simulations on the single thimble ${\cal J}_{\sigma_{0}}$, 
which is expected to dominate the path-integral.
We have measured the average residual phase, number density and scalar condensate.
The average residual phase almost always stays greater than $0.8$  and 
the reweighting works in this model for our choice of the parameter sets.
By comparing our numerical results with the exact ones, 
we have examined to what extent the HMC method works and the single thimble 
${\cal J}_{\sigma_{0}}$ reproduces the exact result.

%
%

%
The numerical results of the number density and scalar condensate reproduce the exact ones 
at small $L\simeq 4, 8$ and large $\beta \simeq 3, 6$.  We also observed that these numerical results scale toward the continuum limit keeping $L (ma)$ and $\beta (ma)$ fixed. These results imply 
that the single-thimble approximation with ${\cal J}_{\sigma_{0}}$ would work 
in the weak coupling region of $ g^2 / m \le 1/6$ 
 and/or 
in the high temperature region of $T/m \ge 1/8$. 

However, we observed the discrepancy in the crossover region 
for smaller $\beta$ and/or larger $L$. 
It persists in the continuum limit at a fixed temperature and
becomes more significant toward the large $L$ limit,  or the low-temperature limit.
These numerical results are quite consistent with our analytical study of  
the model\cite{Fujii:2015bua}. 
Our studies clearly show that the contributions of subdominant thimbles should be summed up 
in order to reproduce the rapid crossover and the first-order transition 
in the low-temperature limit. 

In the Monte Carlo methods formulated on the Lefschetz thimbles, 
it is not straightforward to sum up the contributions over the set of the relevant thimbles.
This is because one need to obtain the relative (complex) weight factors
$\{ {\rm e}^{- S[\sigma]}  \, Z_{\sigma} \}$ (See Eqs.~(\ref{eq:partion-function-by-thimbles})). 
However,  a general method to compute these quantities is not known so far. 
It is then highly desirable to 
devise an efficient way to perform the multi-thimble integration by extending the Monte Carlo algorithms for practical applications of the Lefschetz thimble integration to fermionic systems 
with the sign problem.

\acknowledgments
When we were finishing this and the related articles, we were informed by Y.~Hidaka that
they have obtained the similar result about the multi-thimble contributions necessary to reproduce the non-analytic behavior of observables in the one-site Hubberd model\cite{Tanizaki:2015rda}. 
We would like to thank him for sharing their result with us.
H.F. acknowledges a userful conversation with Y. Tanizaki on this and the related works.
We are grateful to D.~Kadoh for allowing us to use his numerical codes.
This work is supported in part by JSPS KAKENHI Grant Numbers 24540255 (H.F.),  24540253 (Y.K.). S.K.\ is supported by the Advanced Science Measurement Research Center at Rikkyo University.


\begin{thebibliography}{99}




\bibitem{deForcrand:2010ys} 
  P.~de Forcrand,
  PoS LAT {\bf 2009}, 010 (2009)
  [arXiv:1005.0539 [hep-lat]].



\bibitem{Parisi:1984cs} 
  G.~Parisi,
  Phys.\ Lett.\ B {\bf 131}, 393 (1983).

\bibitem{Klauder:1983zm} 
  J.~R.~Klauder,
  J.\ Phys.\ A {\bf 16}, L317 (1983).

\bibitem{Klauder:1983sp} 
  J.~R.~Klauder,
  Phys.\ Rev.\ A {\bf 29}, 2036 (1984).

%
%
%

%
%
%


%

\bibitem{Witten:2010cx} 
  E.~Witten,
  AMS/IP Stud.\ Adv.\ Math.\  {\bf 50}, 347 (2011)
  [arXiv:1001.2933 [hep-th]].

\bibitem{Witten:2010zr} 
  E.~Witten,
  arXiv:1009.6032 [hep-th].

\bibitem{Pham:1983} 
 F.~Pham,
 Proc. Symp. in Pure Math 40, part 2 (1983).



\bibitem{Aarts:2008rr} 
  G.~Aarts and I.~-O.~Stamatescu,
  JHEP {\bf 0809}, 018 (2008)
  [arXiv:0807.1597 [hep-lat]].
  
\bibitem{Aarts:2008wh} 
  G.~Aarts,
  Phys.\ Rev.\ Lett.\  {\bf 102}, 131601 (2009)
  [arXiv:0810.2089 [hep-lat]].

\bibitem{Aarts:2009hn} 
  G.~Aarts,
  JHEP {\bf 0905}, 052 (2009)
  [arXiv:0902.4686 [hep-lat]].
 
\bibitem{Aarts:2009yj} 
  G.~Aarts,
  PoS LAT {\bf 2009}, 024 (2009)
  [arXiv:0910.3772 [hep-lat]].

\bibitem{Aarts:2009dg} 
  G.~Aarts, F.~A.~James, E.~Seiler and I.~-O.~Stamatescu,
  Phys.\ Lett.\ B {\bf 687}, 154 (2010)
  [arXiv:0912.0617 [hep-lat]].
  
\bibitem{Aarts:2009uq} 
  G.~Aarts, E.~Seiler and I.~-O.~Stamatescu,
  Phys.\ Rev.\ D {\bf 81}, 054508 (2010)
  [arXiv:0912.3360 [hep-lat]].

\bibitem{Aarts:2010aq} 
  G.~Aarts and F.~A.~James,
  JHEP {\bf 1008}, 020 (2010)
  [arXiv:1005.3468 [hep-lat]].

\bibitem{Aarts:2010gr} 
  G.~Aarts and K.~Splittorff,
  JHEP {\bf 1008}, 017 (2010)
  [arXiv:1006.0332 [hep-lat]].

\bibitem{Aarts:2011ax} 
  G.~Aarts, F.~A.~James, E.~Seiler and I.~-O.~Stamatescu,
  Eur.\ Phys.\ J.\ C {\bf 71}, 1756 (2011)
  [arXiv:1101.3270 [hep-lat]].
  
\bibitem{Aarts:2011sf} 
  G.~Aarts, F.~A.~James, E.~Seiler and I.~O.~Stamatescu,
  PoS LATTICE {\bf 2011}, 197 (2011)
  [arXiv:1110.5749 [hep-lat]].

\bibitem{Aarts:2011zn} 
  G.~Aarts and F.~A.~James,
  JHEP {\bf 1201}, 118 (2012)
  [arXiv:1112.4655 [hep-lat]].
  
%

\bibitem{Seiler:2012wz} 
  E.~Seiler, D.~Sexty and I.~-O.~Stamatescu,
  Phys.\ Lett.\ B {\bf 723}, 213 (2013)
  [arXiv:1211.3709 [hep-lat]].
  
\bibitem{Aarts:2012ft} 
  G.~Aarts, F.~A.~James, J.~M.~Pawlowski, E.~Seiler, D.~Sexty and I.~O.~Stamatescu,
  JHEP {\bf 1303}, 073 (2013)
  [arXiv:1212.5231 [hep-lat]].
 
\bibitem{Pawlowski:2013pje} 
  J.~M.~Pawlowski and C.~Zielinski,
  Phys.\ Rev.\ D {\bf 87}, 094503 (2013)
  [arXiv:1302.1622 [hep-lat]].

\bibitem{Pawlowski:2013gag} 
  J.~M.~Pawlowski and C.~Zielinski,
  Phys.\ Rev.\ D {\bf 87}, 094509 (2013)
  [arXiv:1302.2249 [hep-lat]].
  
\bibitem{Aarts:2013bla} 
  G.~Aarts,
  PoS LATTICE {\bf 2012}, 017 (2012)
  [arXiv:1302.3028 [hep-lat]].

\bibitem{Aarts:2013uxa} 
  G.~Aarts, L.~Bongiovanni, E.~Seiler, D.~Sexty and I.~-O.~Stamatescu,
  arXiv:1303.6425 [hep-lat].
  
\bibitem{Aarts:2013uza} 
  G.~Aarts, P.~Giudice and E.~Seiler,
  Annals Phys.\  {\bf 337}, 238 (2013)
  [arXiv:1306.3075 [hep-lat]].
  
\bibitem{Sexty:2013ica} 
  D.~Sexty,
  ``Simulating full QCD at nonzero density using the complex Langevin equation,''
  arXiv:1307.7748 [hep-lat].

%
%
\bibitem{Aarts:2013fpa} 
  G.~Aarts,
  Phys.\ Rev.\ D {\bf 88}, no. 9, 094501 (2013)
  [arXiv:1308.4811 [hep-lat]].

  
\bibitem{Giudice:2013eva} 
  P.~Giudice, G.~Aarts and E.~Seiler,
  ``Localised distributions in complex Langevin dynamics,''
  arXiv:1309.3191 [hep-lat].

\bibitem{Mollgaard:2013qra} 
  A.~Mollgaard and K.~Splittorff,
  Phys.\ Rev.\ D {\bf 88}, no. 11, 116007 (2013)
  [arXiv:1309.4335 [hep-lat]].
  
\bibitem{Sexty:2013fpt} 
  D.~Sexty,
  PoS LATTICE {\bf 2013}, 199 (2014)
  [arXiv:1310.6186 [hep-lat]].
  

\bibitem{Aarts:2013nja} 
  G.~Aarts, L.~Bongiovanni, E.~Seiler, D.~Sexty and I.~O.~Stamatescu,
  PoS LATTICE {\bf 2013}, 451 (2014)
  [arXiv:1310.7412 [hep-lat]].

\bibitem{Bongiovanni:2013nxa} 
  L.~Bongiovanni, G.~Aarts, E.~Seiler, D.~Sexty and I.~O.~Stamatescu,
  PoS LATTICE {\bf 2013}, 449 (2014)
  [arXiv:1311.1056 [hep-lat]].
 
 
\bibitem{Aarts:2014nxa} 
  G.~Aarts, L.~Bongiovanni, E.~Seiler and D.~Sexty,
  JHEP {\bf 1410}, 159 (2014)
  [arXiv:1407.2090 [hep-lat]].
  
\bibitem{Aarts:2014bwa} 
  G.~Aarts, E.~Seiler, D.~Sexty and I.~O.~Stamatescu,
  Phys.\ Rev.\ D {\bf 90}, no. 11, 114505 (2014)
  [arXiv:1408.3770 [hep-lat]].
  
\bibitem{Sexty:2014zya} 
  D.~Sexty,
  Nucl.\ Phys.\ A {\bf 931}, 856 (2014)
  [arXiv:1408.6767 [hep-lat]].
    
\bibitem{Sexty:2014dxa} 
  D.~Sexty,
  PoS LATTICE {\bf 2014}, 016 (2014)
  [arXiv:1410.8813 [hep-lat]].
  
\bibitem{Bongiovanni:2014rna} 
  L.~Bongiovanni, G.~Aarts, E.~Seiler and D.~Sexty,
  PoS LATTICE {\bf 2014}, 199 (2014)
  [arXiv:1411.0949 [hep-lat]].
   
\bibitem{Aarts:2014kja} 
  G.~Aarts, F.~Attanasio, B.~J?ger, E.~Seiler, D.~Sexty and I.~O.~Stamatescu,
  PoS LATTICE {\bf 2014}, 200 (2014)
  [arXiv:1411.2632 [hep-lat]].

\bibitem{Aarts:2014cua} 
  G.~Aarts, B.~J?ger, E.~Seiler, D.~Sexty and I.~O.~Stamatescu,
  PoS LATTICE {\bf 2014}, 207 (2014)
  [arXiv:1412.5775 [hep-lat]].
    
\bibitem{Aarts:2014fsa} 
  G.~Aarts, F.~Attanasio, B.~J?ger, E.~Seiler, D.~Sexty and I.~O.~Stamatescu,
  arXiv:1412.0847 [hep-lat].
 
\bibitem{Mollgaard:2014mga} 
  A.~Mollgaard and K.~Splittorff,
  Phys.\ Rev.\ D {\bf 91}, no. 3, 036007 (2015)
  [arXiv:1412.2729 [hep-lat]].
     
\bibitem{Makino:2015ooa} 
  H.~Makino, H.~Suzuki and D.~Takeda,
  arXiv:1503.00417 [hep-lat].
   
\bibitem{Aarts:2015oka} 
  G.~Aarts, E.~Seiler, D.~Sexty and I.-O.~Stamatescu,
  arXiv:1503.08813 [hep-lat].
    
\bibitem{Nishimura:2015pba} 
  J.~Nishimura and S.~Shimasaki,
  Phys.\ Rev.\ D {\bf 92}, no. 1, 011501 (2015)
  [arXiv:1504.08359 [hep-lat]].

   
\bibitem{Aarts:2015yba} 
  G.~Aarts, F.~Attanasio, B.~J?ger, E.~Seiler, D.~Sexty and I.~O.~Stamatescu,
  Acta Phys.\ Polon.\ Supp.\  {\bf 8}, no. 2, 405 (2015)
  [arXiv:1506.02547 [hep-lat]].

\bibitem{Nagata:2015uga} 
  K.~Nagata, J.~Nishimura and S.~Shimasaki,
  arXiv:1508.02377 [hep-lat].
  
\bibitem{Fodor:2015doa} 
  Z.~Fodor, S.~D.~Katz, D.~Sexty and C.~T?r?k,
  arXiv:1508.05260 [hep-lat].
  
\bibitem{Tsutsui:2015tua} 
  S.~Tsutsui and T.~M.~Doi,
  arXiv:1508.04231 [hep-lat].




\bibitem{Cristoforetti:2012su} 
  M.~Cristoforetti {\it et al.}  [AuroraScience Collaboration],
  Phys.\ Rev.\ D {\bf 86}, 074506 (2012)
  [arXiv:1205.3996 [hep-lat]].

\bibitem{Cristoforetti:2013wha} 
  M.~Cristoforetti, F.~Di Renzo, A.~Mukherjee and L.~Scorzato,
  Phys.\ Rev.\ D {\bf 88}, no. 5, 051501 (2013)
  [arXiv:1303.7204 [hep-lat]].


\bibitem{Fujii:2013sra} 
  H.~Fujii, D.~Honda, M.~Kato, Y.~Kikukawa, S.~Komatsu and T.~Sano,
  JHEP {\bf 1310}, 147 (2013)
  [arXiv:1309.4371 [hep-lat]].




\bibitem{Mukherjee:2014hsa} 
  A.~Mukherjee and M.~Cristoforetti,
  Phys.\ Rev.\ B {\bf 90}, no. 3, 035134 (2014)
  [arXiv:1403.5680 [cond-mat.str-el]].

\bibitem{DiRenzo:2015foa} 
  F.~Di Renzo and G.~Eruzzi,
  arXiv:1507.03858 [hep-lat].

\bibitem{Tanizaki:2014tua} 
  Y.~Tanizaki,
  Phys.\ Rev.\ D {\bf 91}, no. 3, 036002 (2015)
  [arXiv:1412.1891 [hep-th]].

\bibitem{Kanazawa:2014qma} 
  T.~Kanazawa and Y.~Tanizaki,
  JHEP {\bf 1503}, 044 (2015)
  [arXiv:1412.2802 [hep-th]].

\bibitem{Cristoforetti:2014gsa} 
  M.~Cristoforetti, F.~Di Renzo, G.~Eruzzi, A.~Mukherjee, C.~Schmidt, L.~Scorzato and C.~Torrero,
  Phys.\ Rev.\ D {\bf 89}, no. 11, 114505 (2014)
  [arXiv:1403.5637 [hep-lat]].

\bibitem{Tanizaki:2014xba} 
  Y.~Tanizaki and T.~Koike,
  Annals Phys.\  {\bf 351}, 250 (2014)
  [arXiv:1406.2386 [math-ph]].


\bibitem{Tanizaki:2015pua} 
  Y.~Tanizaki, H.~Nishimura and K.~Kashiwa,
  Phys.\ Rev.\ D {\bf 91}, no. 10, 101701 (2015)
  [arXiv:1504.02979 [hep-th]].

\bibitem{Cherman:2014ofa} 
  A.~Cherman, D.~Dorigoni and M.~Unsal,
  arXiv:1403.1277 [hep-th].

\bibitem{Behtash:2015kna} 
  A.~Behtash, T.~Sulejmanpasic, T.~Schaefer and M.~Unsal,
  arXiv:1502.06624 [hep-th].

\bibitem{Scorzato:2015}
L.~Scorzato, ``The Lefschetz thimble and the sign problem'', plenary talk at Lattice 2015. 

\bibitem{Fukushima:2015qza} 
  K.~Fukushima and Y.~Tanizaki,
  arXiv:1507.07351 [hep-th].

\bibitem{Tanizaki:2015rda} 
  Y.~Tanizaki, Y.~Hidaka and T.~Hayata,
  arXiv:1509.07146 [hep-th].

\bibitem{Fujii:2015bua} 
  H.~Fujii, S.~Kamata and Y.~Kikukawa,
  arXiv:1509.08176 [hep-lat].



\bibitem{Pawlowski:2014ada} 
  J.~M.~Pawlowski, I.~O.~Stamatescu and C.~Zielinski,
  arXiv:1402.6042 [hep-lat].

\bibitem{Kogut:1974ag} 
  J.~B.~Kogut and L.~Susskind,
  Phys.\ Rev.\ D {\bf 11}, 395 (1975).


\bibitem{Hasenfratz:1983ba} 
  P.~Hasenfratz and F.~Karsch,
  Phys.\ Lett.\ B {\bf 125}, 308 (1983).

 
\bibitem{Cohen:2003kd} 
  T.~D.~Cohen,
  Phys.\ Rev.\ Lett.\  {\bf 91}, 222001 (2003)
  [hep-ph/0307089].


%
%
%
%
%
%
%
%
%
%
%

%
%
%


%

  
\bibitem{Mol:2008}
 L.G. Molinari, Linear Algebra and its Applications 429, 2221-2226 (2008)

\bibitem{Hahna:2006} 
  T.~Hahna,
arXiv:physics/0607103 [physics.comp-ph].




  
\end{thebibliography}
\end{document}